\documentclass[preprint,showpacs,preprintnumbers,superscriptaddress,aps]{revtex4}
\makeatletter
\renewcommand*{\@fnsymbol}[1]{\@alph{#1})}
\makeatother
\usepackage[colorlinks,linkcolor=blue,anchorcolor=blue,citecolor=blue]{hyperref}
\usepackage{amsmath}
\usepackage{amssymb}
\usepackage{amsfonts}
\usepackage{makecell} 
\usepackage{graphicx}
\usepackage{dcolumn}
\usepackage{bm}

\makeatletter
\def\@pacs@print{} 
\def\pacs#1{\def\@pacs{#1}} 
\makeatother
\begin{document}

\title{On the degradation of hot spot performance due to mid-to-high-mode hydrodynamic instabilities} 


\small
\author{Dongxue Liu}
\affiliation{Shanghai Institute of Laser Plasma, Shanghai 201800, China
}
\author{Jiaqin Dong}
\email[]{Authors to whom correspondence should be addressed: dongjiaqin@hotmail.com and jzheng@ustc.edu.cn}
\affiliation{Shanghai Institute of Laser Plasma, Shanghai 201800, China
}
\author{Yunxing Liu}
\affiliation{Shanghai Institute of Laser Plasma, Shanghai 201800, China
}
\author{Zhiyu He}
\affiliation{Shanghai Institute of Laser Plasma, Shanghai 201800, China
}
\author{Wei Wang}
\affiliation{Shanghai Institute of Laser Plasma, Shanghai 201800, China
}
\author{Jinren Sun}
\affiliation{Shanghai Institute of Laser Plasma, Shanghai 201800, China
}
\author{Yuqiu Gu}
\affiliation{Shanghai Institute of Laser Plasma, Shanghai 201800, China
}
\author{Xiuguang Huang}
\affiliation{Shanghai Institute of Laser Plasma, Shanghai 201800, China
}




\author{Jian Zheng}
\email[]{Authors to whom correspondence should be addressed: dongjiaqin@hotmail.com and jzheng@ustc.edu.cn}
\affiliation{Department of Plasma Physics and Fusion Engineering, and CAS Key Laboratory of Frontier Physics in Controlled Nuclear Fusion, University of Science and Technology of China, Hefei 230026, China
}
\affiliation{Collaborative Innovation Center of IFSA, Shanghai Jiao Tong University, Shanghai, 200240, China}

\begin{abstract}
In an ignited design of inertial confinement fusion, the role of mid‑to-high-mode hydrodynamic instabilities in degrading hot-spot performance, beyond reducing temperature, remains unclear. To address this, we propose an isobaric criterion to assess the isobaric assumption that forms the theoretical basis of the hot spot. The most dangerous mode $l=12$ is determined through a balance between perturbation growth and ablation stabilization induced by thermal conduction. Thermal conduction outperforms convection when the Peclet number is much less than 1. Therefore, for mid‑to‑high modes, thermal conduction makes the hot spot isobaric before the outer mass inflow restores the lost heat. Consequently, neglecting thermal conduction overestimates pressure and underestimates volume. These results enhance our understanding of mid-to-high modes in degrading hot-spot performance, and suggest that thermal conduction losses may reduce performance even if perturbations are nearly stabilized by ablation.

\hspace{-10pt}Keywords: hydrodynamic instabilities, hot spot, isobaric assumption, thermal conduction
\end{abstract}


\maketitle 

\section{Introduction}
The National Ignition Facility (NIF) has achieved ignition \cite{abu2022lawson,abu2024achievement,lindl2026key} in indirect‑drive inertial confinement fusion (ICF). In this scheme, radiation ablation induces the rocket effect, compressing the shell, heating the central low‑density gas, and forming a hot spot \cite{lindl1995development,lindl2004physics}. However, the ignition possibility is challenged by $P_1$
asymmetry \cite{spears2014mode,rinderknecht2020azimuthal}, $P_2$ asymmetry \cite{ralph2024impact,divol2024thermonuclear}, and CH‑DT mixing \cite{regan2013hot,bachmann2020localized,bachmann2022measurement}, all of which significantly degrade the hot‑spot quality. In the pursuit of high gain, direct drive remains attractive due to its higher theoretical gain and simpler target design \cite{RN506}. However, in addition to the aforementioned low‑mode asymmetries and mixing, laser irradiation introduces further instabilities in direct drive, including low-mode asymmetries ($2<l\leq6$) \cite{PhysRevLett.120.125001,PhysRevLett.127.075001}, mid-mode asymmetries ($l>6$) \cite{patel2017signatures,baltazar2022diagnosing,RN413} and high-mode nonuniformities ($l\geq 20$)\cite{patel2023effects,michel2017measurement}. How these instabilities degrade the hot‑spot quality remains poorly understood, hindering the assessment and optimization of the hot‑spot performance.

The hot‑spot performance is usually described by its power-balance relation:
\begin{equation}
 \frac{dE_{I}}{dt} = W_{\alpha} + W_{PdV}-W_{rad}-W_{cond}.  
\end{equation}
Here, $E_I$ is the hot‑spot internal energy; $W_\alpha$, $W_{PdV}$, $W_{rad}$, and $W_{cond}$ are the power terms for $\alpha$-self‑heating, compression work, bremsstrahlung radiation, and thermal conduction loss, respectively. Because the flow velocity is much less than the sound speed, the hot spot remains nearly isobaric. Additionally, $W_{cond}$ and low-energy $W_{rad}$ can be compensated by an increase in hot‑spot mass. These two features make the isobaric \cite{atzeni1984inertial,meyer1982energy} and adiabatic \cite{woo2018effects,kritcher2022design} assumptions a common choice for modeling hot‑spot formation when $\alpha$ self-heating is weak. By adjusting three quantities, namely, adiabat of the stagnated fuel, hydrodynamic efficiency, and spark radius, the isobaric model obtains the gain curve, which precisely matches the LLNL results \cite{meyer1982energy,RN973} while overestimates the effect of preheat on implosion performance \cite{piriz1992energy}. To the best of our knowledge, no study has yet been reported on the influence of hydrodynamic instabilities on the isobaric assumption, especially for mid‑to‑high modes. 

Models based on isobaric or isobaric‑adiabatic assumptions have been developed to study the effects of hydrodynamic instability. Under the isobaric and adiabatic assumption, Hurricane et al. quantified low‑mode asymmetries via the areal‑density weighted harmonic mean (WHM) \cite{hurricane2020analytic,hurricane2022extensions,hurricane2025present}. Under the isobaric assumption, the three‑dimensional piston model incorporates $W_{rad}$ and $W_{cond}$ to evaluate engineering features such as support tents \cite{springer20193d}. However, mid‑to‑high mode instabilities remain poorly described by these models, largely because the underlying degradation mechanisms are not fully understood. Simple treatments that equate such instabilities with a reduction in the clean hot‑spot volume tend to overestimate their impact \cite{chang2010generalized}. Compared to P1 asymmetry \cite{spears2014mode,gu2014new} converting compressional work into residual kinetic energy (RKE), higher‑mode instabilities ($l>6$) \cite{bose2017physics} degrade implosion performance through two pathways. First, increasing the effective hot‑spot surface area enhances \(W_{cond}\), Second, nonsimultaneous stagnation or reduced shell compressibility reduces $W_{PdV}$ to heat the hot spot.

In this paper, we propose an isobaric criterion, $\frac{\rho R_{\max}}{\rho R_{{ideal}}} =\frac{\Delta R_{{shell}}}{\Delta R_{{dece.}}}$, by linking the hot-spot isobaric assumption to hydrodynamic instabilities. The isobaric assumption holds if perturbations in areal density $\frac{\rho R_{\max}}{\rho R_{{ideal}}}$ does not exceed the ratio of the stagnated shell thickness $\Delta R_{{shell}}$ to coasting distance \(\Delta R_{{dece.}}\). In an ignited design \cite{kritcher2022design,kritcher2024design}, the most dangerous mode $l=12$ is determined via a balance between perturbation growth and ablation stabilization induced by thermal conduction. Thermal conduction outperforms convection when Peclet number is much less than unity. Therefore, for mid‑to‑high modes, thermal conduction makes the hot spot isobaric before the outer mass inflow restores the lost heat. Consequently, compared with the case of only mechanical compression, pressure is overestimated and volume is underestimated. These results enhance our understanding of mid-to-high modes in degrading hot-spot performance, and suggest that $W_{cond}$ may reduce performance even if the perturbations are nearly stabilized by ablation.

The paper is organized as follows. In Section II, the isobaric assumption is proposed and validated by two‑dimensional radiation-hydrodynamic simulations \cite{fryxell2000flash}. In Section III, we propose the most dangerous mode and analyze the physical mechanisms of mid-to-high mode instabilities degrading hot-spot performance. In Section VI, we draw our conclusions.

\section{Impact of Hydrodynamic Instabilities on Isobaric assumption}
Hot spot characteristics at stagnation, including minimum volume, peak internal energy and maximum pressure, are general metrics for evaluating various implosion designs. Understanding how hydrodynamic instabilities affect these metrics is essential for uncovering the physical mechanisms of hot spot degradation. Previous studies have commonly used the isobaric and adiabatic model. Adiabaticity assumes that the hot spot is large enough with no heat escape. In this section, we propose a criteria for assessing the validity of the isobaric assumption under the influence of hydrodynamic instabilities. This provides the conditions for adopting the isobaric and adiabatic model as a simplified theoretical baseline.

\subsection{Two-dimensional simulation settings}
We use the radiation‑hydrodynamic code FLASH \cite{fryxell2000flash} to simulate hot‑spot formation in two‑dimensional (2D) cylindrical geometry. The initial flow field is chosen based on two considerations. First, since hot spot formation in central ignition is essentially identical in various drive schemes, we adopt the initial flow field from an ignited design. Second, this design has only 4\% remaining ablator mass, resulting in a thin shell and high implosion velocity. This make the design sensitive to hydrodynamic instabilities, which may exacerbated by three-dimensional (3D) effects \cite{liu2025three,PhysRevLett.120.125001}. To avoid 3D effects while maximizing rocket efficiency, which is necessary for high-gain direct‑drive, we artificially reduce the number of radiation energy groups and increase the remaining ablator mass fraction to 20\%. This reduces the ablated mass, but does not reduce the ratio of the shell kinetic energy to $W_{I}$. Therefore, we can study the time evolution of multi‑scale instabilities and its  effects on the isobaric assumption for ignition‑relevant hot spots.
\begin{figure}[h]
\begin{minipage}{0.7\textwidth}
\includegraphics[width=0.93\linewidth]{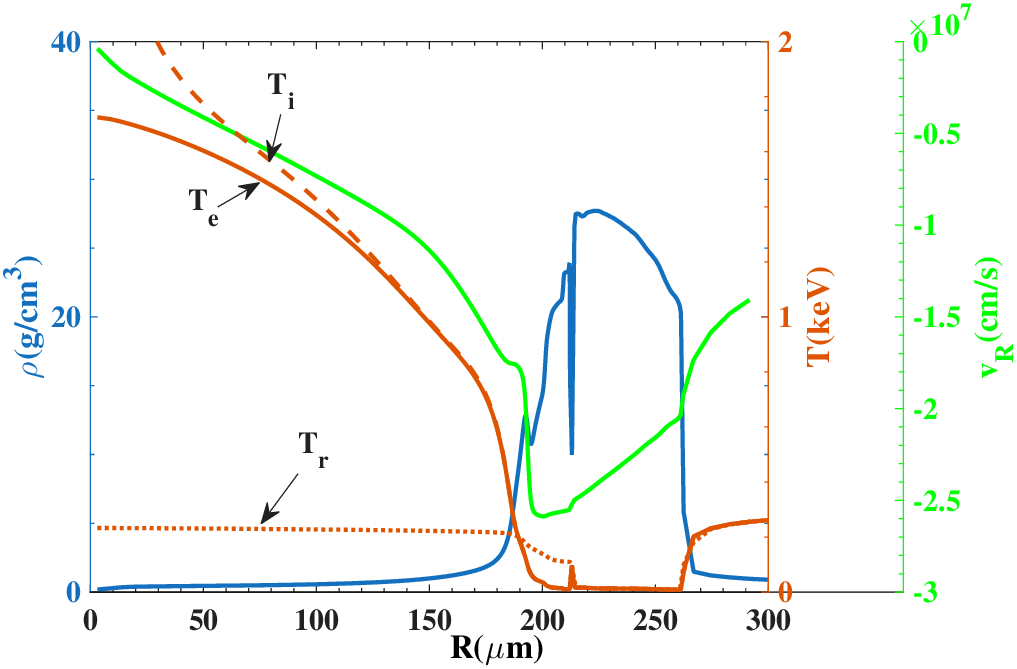}%
 \end{minipage}
 \caption{\label{FIG. 1} Two‑dimensional simulation settings, Equilibrium flow field distribution from 1D MULTI in spherical coordination \cite{ramis1988multi}. The simulation settings of MULTI are detailed in the Appendix.}
 \end{figure}

The two‑dimensional cylindrical simulation begins at the onset of the deceleration phase, when the main shock rebounds and reaches the cold shell's inner surface. The initial equilibrium flow fields, including density $\rho$, electron temperature $T_e$, ion temperature $T_i$, radiation temperature $T_r$, and velocity $v_R$ are obtained from the one‑dimensional MULTI \cite{ramis1988multi} code. These profiles in spherical geometry are displayed in Fig. \ref{FIG. 1}(a) and are converted to cylindrical coordinates in 2D FLASH simulations. 
The high-density carbon (HDC) ablator and DT fuel are described by the Sesame equation of state \cite{mchardy2018introduction}. The Rosseland and Planck mean opacities are adopted for the radiation transport. The radiation transport is treated with a 20-group approximation. The flux-limiter is 0.06. The simulation domain is \(R = [0\ \mu{m}, 300\ \mu{m}]\), \(Z = [-300\ \mu{m}, 300\ \mu{m}]\), with a resolution ranging from \(0.78\ \mu{m}\) to \(0.06\ \mu{m}\).
 \subsection{The Isobaric and Adiabatic Hot Spot}
In an ideal spherical implosion, 
we clarify the definition and the isobaric-adiabatic characteristics of the hot spot, laying a theoretical foundation for linking the isobaric assumption with hydrodynamic instabilities at interfaces. 
\begin{figure}[h] 
 \begin{minipage}{0.6\textwidth}
(a)\includegraphics[width=0.95\linewidth]{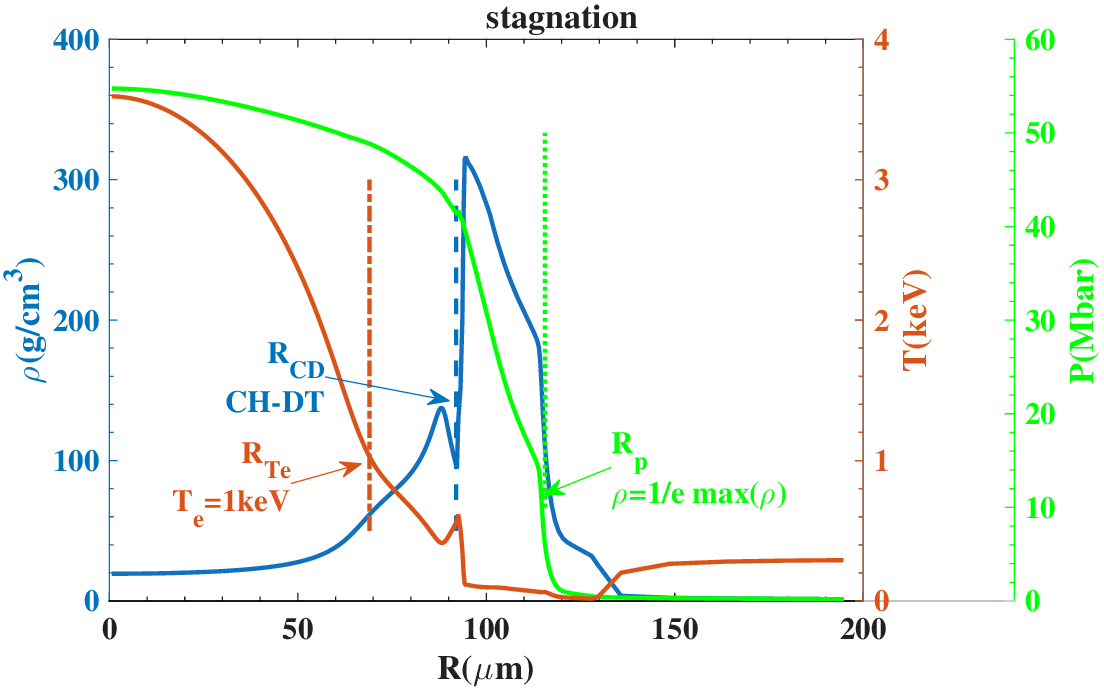}%
\end{minipage}
\begin{minipage}{0.35\textwidth}
(b)\includegraphics[width=0.95\linewidth]{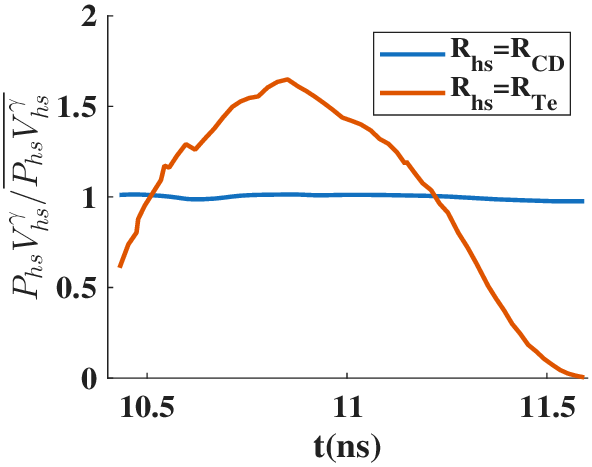}%
\end{minipage}
 \caption{\label{FIG. 11} Ideal spherical implosion. (a) Profiles of temperature, density and pressure at stagnation. Three vertical lines (from left to right) mark the isothermal surface $T_e = 1$keV ($R_{Te}$), the CH‑DT interface ($R_{CD}$), and the outer interface ($R_p$) where the density equals the $\frac{1}{e}$ of its maximum. $R_p$ corresponds to the reflective shock front and serves as the isobaric surface. For designs where the remaining ablator mass is smaller, $R_p$ could exceed $R_{CD}$. (b) Normalized adiabaticity, where $P_{hs}=\int_0^{V_{hs}}PdV/(V_{hs})$ is the hot-spot pressure, $V_{hs}=4\pi/3R_{hs}^3$ is the hot-spot volume and $\overline{\bullet}$ represent time average during the deceleration phase. $\frac{P_{hs}V_{hs}^{\gamma}}{\overline{P_{hs}V_{hs}^{\gamma}}}=1$ means adiabaticity during the deceleration phase.} 
 \end{figure}

A cold, dense shell surrounds the central hot, low‑density gas. At stagnation, the volume of the central gas is minimized, its internal energy is maximized, and the hot spot forms. Three characteristic interfaces during this process are shown in Fig. \ref{FIG. 11}(a). From the center outward, they are the surface where $T_e=1$ keV ($R_{Te}$), the CH‑DT interface ($R_{CD}$), and the surface where  $\rho=\max(\rho)/{e}$ ($R_{p}$). $R_{Te}$ can be adopted as the boundary of the hot spot \cite{woo2018effects}. $R_{CD}$ marks the abrupt change in radiation opacity. $R_{p}$ corresponds to the leading edge of the reflective shock and can be regarded as the isobaric boundary, because pressure remains nearly constant within it. 

Unlike the isobaric assumption, the adiabatic assumption has different implications depending on the hot-spot definition. As shown in Fig. \ref{FIG. 11}(b), if the boundary of the hot spot is defined as $R_{hs}=R_{Te}$, its adiabatic assumption requires that $P_{hs}V_{hs}^{\gamma}$ remains invariant at stagnation not throughout the entire deceleration phase. Here, $P_{hs}=\int_0^{V_{hs}}PdV/(V_{hs})$ is the hot-spot pressure, $V=4\pi/3R_{hs}^3$ is the hot-spot volume. If $R_{hs}=R_{DT}$, the hot spot remains adiabatic throughout the entire deceleration phase, i.e. $\frac{P_{hs}V_{hs}^{\gamma}}{\overline{P_{hs}V_{hs}^{\gamma}}}=1$, where $\overline{\bullet}$ represents time average during the deceleration phase. 
\subsection{isobaric criterion for the hot spot}
Perturbations at the hot‑spot interface, amplified by RTI, RMI, and BP \cite{bell1954alamos,zhou2025instabilities}, develop into spikes. Once the spike depth exceeds $\Delta R_{sh}$, the external low‑pressure region enters the hot spot, breaking the isobaric assumption. Accordingly, we propose an isobaric criterion, using areal‑density perturbations at stagnation to judge the isobaric assumption.

\begin{figure}[h]
\begin{minipage}{0.6\textwidth}
\includegraphics[width=0.93\linewidth]{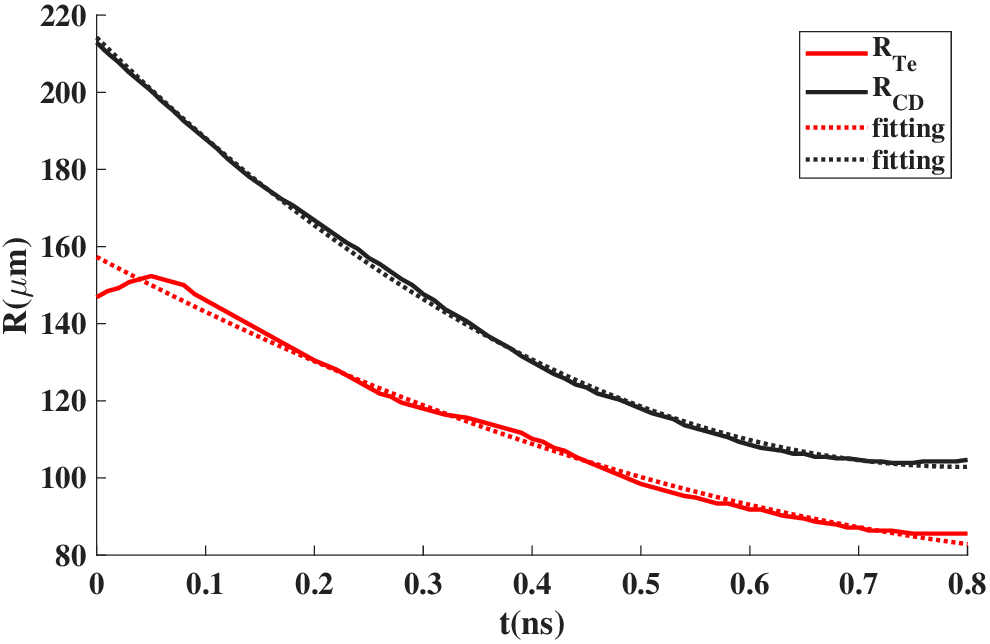}%
 \end{minipage}
 \caption{\label{FIG. 3}Temporal evolution of the interface positions. The interfaces decelerate approximately uniformly, with quadratic fits shown as dotted lines.} 
 \end{figure}
 Assuming constant shell pressure 
 \(P_{sh}\) and areal density \(\rho R\) during deceleration, the cold shell is decelerated by the central hot gas. As shown in Fig. \ref{FIG. 3}, the hot‑spot interface undergoes approximately uniform deceleration regardless of the boundary definition ($R_{hs}=R_{Te}$ or $R_{hs}=R_{CD}$), satisfying
\begin{equation}
v_{imp}^2 = 2 \frac{P_{hs} - P_{sh}}{\rho R_{ideal}} \Delta R_{dece.} \label{1a},
\end{equation}
where \(\Delta R_{dece.}\)
 is the coasting distance, and $v_{imp}$ is the shell implosion velocity at the onset of deceleration. Once the spike depth exceeds the stagnated shell thickness $\Delta R_{sh}$, the external low‑pressure region is adjacent to the hot spot in the non‑radial direction. Non-radial diffusion thereby breaks the isobaric assumption. Conversely, when the spike penetration depth is smaller than
\(\Delta R_{sh}\), the isobaric assumption holds, and the interface in the spike region follows
\begin{equation}
v_{imp}^2 = 2 \frac{P_{hs} - P_{sh}}{\rho R_{\max}} (\Delta R_{dece.} + \Delta R_{sh}) \label{2a}.
\end{equation}
Combining Eq. (\ref{1a}) and Eq.(\ref{2a}) yields the isobaric criterion:
\begin{equation}
\frac{\rho R_{ideal}}{\rho R_{\max}} = \frac{\Delta R_{dece.}}{\Delta R_{dece.} + \Delta R_{sh}}.
\end{equation}
The isobaric assumption holds if $\frac{\rho R_{ideal}}{\rho R_{max}} < \frac{\Delta R_{dece.}}{\Delta R_{dece.} + \Delta R_{sh}}$ and we can use this criterion to perform simulations with isobaric hot spot under the influence of hydrodynamic instabilities.

 \subsection{Temporal evolution of instabilities across scales}
In real implosions, perturbations are localized at specific interfaces. Therefore, we take the perturbed form \(F(R,\theta)=Y_l^0(\theta)\exp(-|R-243.2|/60)\) in simulations with hydrodynamic instabilities. The density perturbation is imposed on the target surface as \(\delta \rho(R,\theta)=\mathcal{A}\rho(R)F(R,\theta)\) to relax the requirement on resolution. Here, $\mathcal{A}$ is a constant and holds the isobaric assumption.

\begin{figure}[h]
\begin{minipage}{0.9\textwidth}
(a)\includegraphics[width=0.95\linewidth]{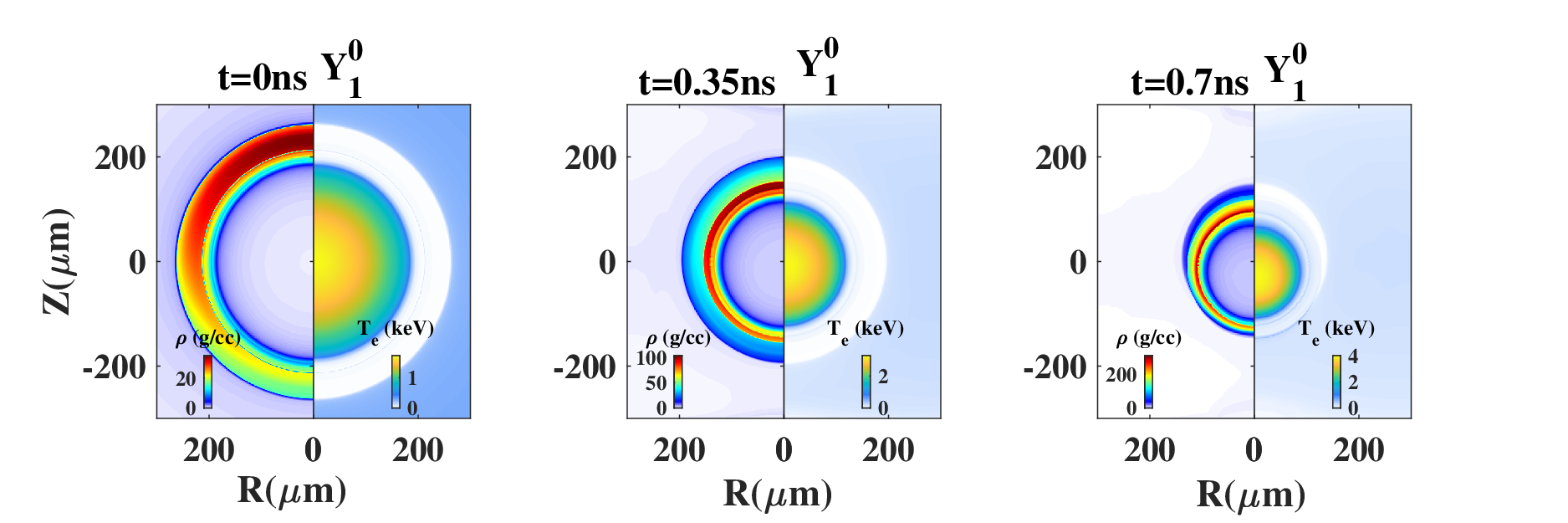}%
 \end{minipage}
 \begin{minipage}{0.9\textwidth}
(b)\includegraphics[width=0.95\linewidth]{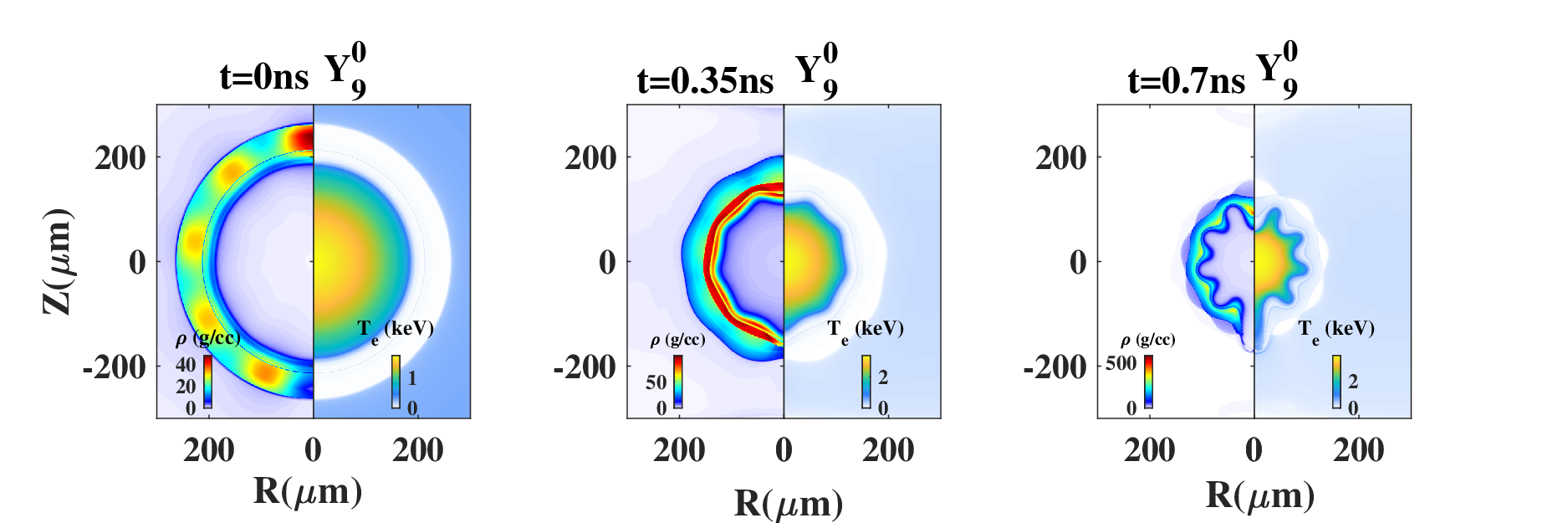}%
 \end{minipage}
  \begin{minipage}{0.9\textwidth}
(c)\includegraphics[width=0.95\linewidth]{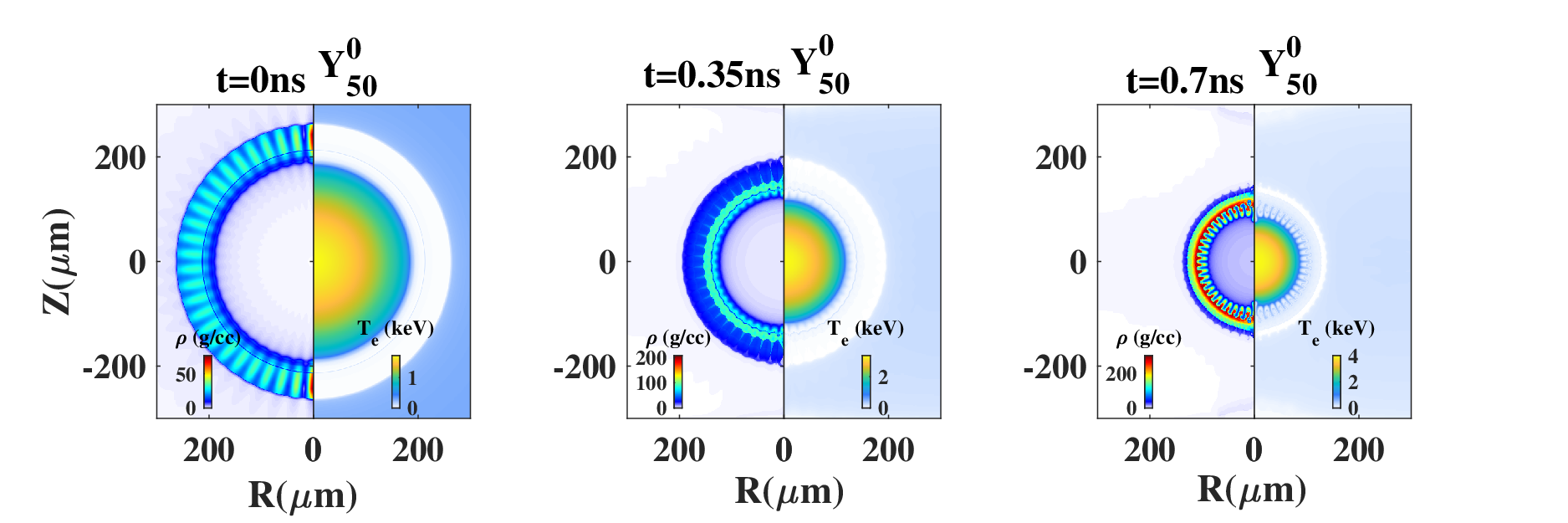}%
 \end{minipage}

 \caption{\label{FIG. 4}
 Density (left) and temperature (right) distributions from the onset of deceleration (t = 0 ns) to stagnation (t = 0.7 ns) under instabilities across scales. (a) $Y_1^0$. (b) $Y_{9}^0$. The anomalously large perturbations in the polar region are R‑Z artifacts and are non‑physical.(c) $Y_{9}^0$.} 
 \end{figure}

 The growth of instabilities across scales is shown in Fig. \ref{FIG. 4}. In regions with high areal density, the deceleration is weak and the flow velocity is high, leading to the formation of spike structures where dense plasma penetrating into low‑density plasma. 
Under \(Y_1^0\) asymmetry in Fig. \ref{FIG. 4}(a), the hot spot deforms and obtains a drift velocity at \(t = 0.7\) ns. In Fig \ref{FIG. 4}(b), $l=9$ mode grow faster due to its shorter wavelength. In contrast, ablation at the surface of the hot spot suppresses the spike growth, so that no significant hot‑spot deformation is observed under \(Y_{50}^0\) at \(t = 0.7\) ns in Fig. \ref{FIG. 4}(c). 

\begin{figure}
     \centering
      \begin{minipage}{0.47\textwidth}
(a)\includegraphics[width=0.95\linewidth]{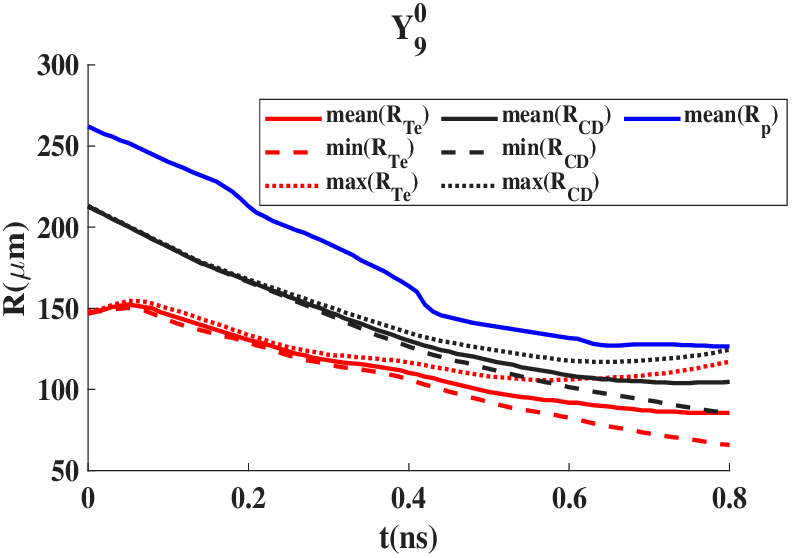}%
 \end{minipage}
 \begin{minipage}{0.47\textwidth}
(b)\includegraphics[width=0.95\linewidth]{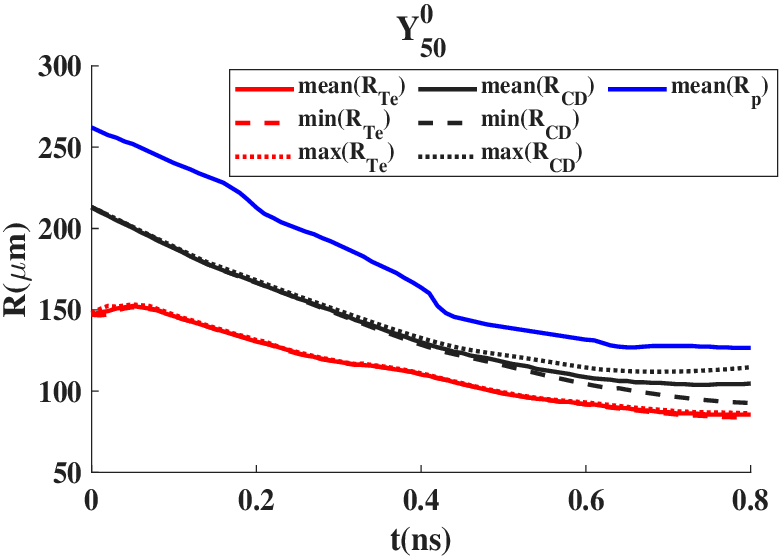}%
 \end{minipage}
     \caption{The temporal evolution of $l=9$ and $l=50$ at $R_{Te}$ and $R_{CD}$. $\min({R_{Te}})$ and $\min({R_{CD}})$ represent the position of spikes, while $\max({R_{Te}})$ and $\max({R_{CD}})$ denote the position of bubbles. Solid lines denote positions from ideal two-dimensional (2D) simulations. Dotted and dashed lines represent the maximum and minimum positions from perturbed 2D simulations.}
     \label{Fig 5}
 \end{figure}

 Furthermore, Fig. \ref{Fig 5}(a) and Fig. \ref{Fig 5}(b) quantify the temporal evolution of $l=9$ and $l=50$ at $R_{Te}$ and $R_{CD}$. We omit the perturbation evolution at \(R_{p}\) because its amplitude, governed solely by the Bell-Plesset (BP) effects \cite{bell1954alamos}, is small, leading to considerable statistical errors.
 For \(Y_9^0\) asymmetry, ablative stabilization is small, because the spikes, denoted by ${mean}(R_{Te})-\min(R_{Te})$ and ${mean}(R_{CD})-\min(R_{CD})$, are nearly equal at $R_{Te}$ and $R_{CD}$.  For \(Y_{50}^0\) asymmetry, ablative stabilization is stronger, as evidenced by the absence of perturbations at \(R_{Te}\) and large perturbation amplitude at \(R_{CD}\), which features a less inverted density profile \cite{hall2024measurement}. The isobaric assumption holds, because \(({mean}(R_{Te}) - \min(R_{Te})) < ({mean}(R_{p}) - {mean}(R_{Te}))\) satisfies for $l=9$ with the largest growth.

\section{The influence of Instabilities}
While low-mode asymmetries degrading hot‑spot performance via $W_{PdV}$ is well understood, the influence of \(W_{{cond}}\) on ignited hot spots remains unclear. \(W_{{cond}}\) couples with \(W_{PdV}\): an increase in \(W_{{cond}}\) reduces residual kinetic energy (RKE), decreasing the surface area of the hot spot and further limiting \(W_{{cond}}\). We propose the most dangerous mode and mid-to-high modes degrading hot-spot pressure via $W_{cond}$. For the first time, The comparsion between a validated isobaric-adiabatic piston model, that includes only \(W_{PdV}\), and simulations is to isolate \(W_{{cond}}\). \(W_{{cond}}\) may reduce performance even if
the perturbations are nearly stabilized by ablation.
\subsection{The most dangerous mode} 
The amplification of perturbations at \(R_{hs}\) is mainly attributed to the combined effects of the BP effect and the RTI. Since BP‑induced growth is mode‑independent, the most dangerous mode is determined by the time integral of the RTI linear growth rate. According to Takabe's formula \cite{RN283,RN286}, 
\[
\gamma = 0.9\sqrt{l/R_{Te}\,g} - 4l/R_{Te}\,v_a,
\] 
two key parameters must be extracted: the ablation velocity \(v_a\) and the acceleration \(g\). 

\begin{figure}[h]
\begin{minipage}{0.47\textwidth}
(a)\includegraphics[width=0.95\linewidth]{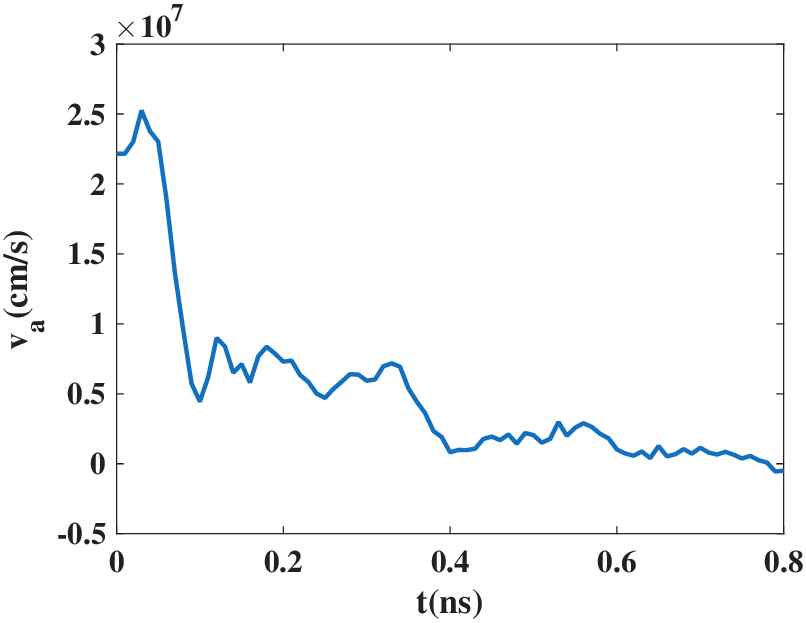}%
\end{minipage}
\begin{minipage}{0.47\textwidth}
(b)\includegraphics[width=0.95\linewidth]{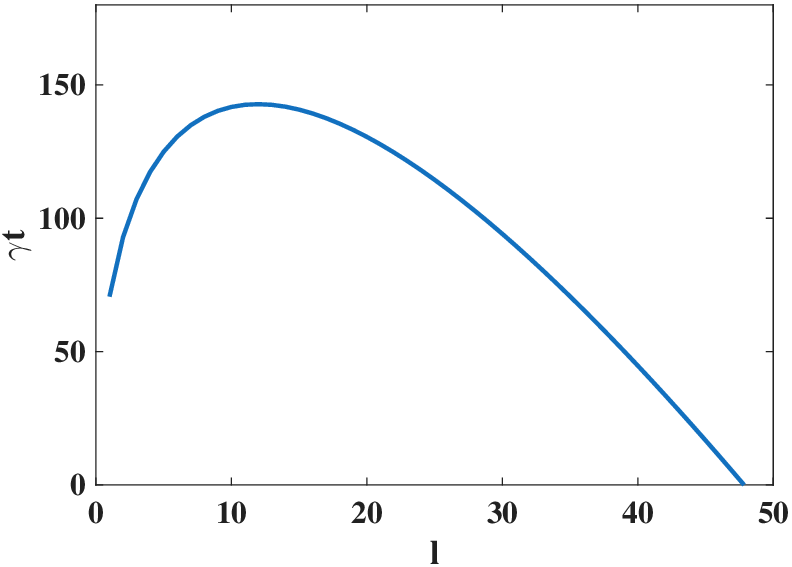}%
\end{minipage}

\caption{\label{FIG. 6} (a) Temporal evolution of ablative velocity. (b) Time-integrated linear growth rate $\gamma t$ of RTI as a function of perturbed mode $l$.} 
\end{figure}

 The ablative velocity \(v_a\) at \(R_{Te}\) is determined from the mass ablation rate \(\dot{m}_a\) as
\begin{equation}
v_a = \frac{\dot{m}_a}{\rho_a},    
\end{equation}
where \(\dot{m}_a = \dfrac{1}{4\pi R_{Te}^2}\dfrac{dm}{dt}\), \(m\) denotes the plasma mass with \(T_e \ge 1\,{keV}\), and \(\rho_a\) is the mass density at \(R_{Te}\). The time history of \(v_a\) is plotted in Fig.~\ref{FIG. 6}(a). It exhibits a stepwise variation, which is mainly attributed to the increase of \(\rho_a\) during compression. To eliminate the influence of compression, we take  \(\overline{v}_a=\int_{0.4}^{0.8}v_adt/0.4=16\,\mu \)m/ns as the average ablative velocity during the deceleration phase.

The acceleration is defined as the second time derivative of  \(R_{T_e}\) and \(g=140.5\,\mu\)m/ns$^2$ can be obtained from the fitting in Fig \ref{FIG. 3}. The time‑integrated linear growth rate \(\gamma t\) as a function of the perturbation mode \(l\) is shown in Fig. \ref{FIG. 6}(b). From this curve, the most dangerous mode is identified as \(l = 12\), and the cutoff mode as \(l = 48\). This is consistent with Fig. \ref{FIG. 4} and Fig. \ref{Fig 5}, where perturbations at $R_{Te}$ increase from $l=1$ to 
$l=9$, while $R_{Te}$ remains almost unperturbed under the influence of $l=50$, indicating that $l=50$ is cut off. Moreover, $l=10$ occurs in Omega's simulations when \(R_b/R_t < 0.85\) \cite{hurricane2023physics}, where $R_b$ and $R_t$ are the radii of the laser beam and the target, respectively. Therefore, understanding mid‑to-high-mode instabilities is critical for assessing and optimizing hot‑spot performance. 
 
\subsection{The isobaric and adiabat piston model}
We use $R_{hs}=R_{CD}$ to ensure the isobaric and adiabatic hot spot. Asymmetries and nonuniformities occur either at the boundary or the interior of the hot spot. The isobaric adiabatic piston model consists of a series of Newton's equations for $N$ piston blocks, i.e., the dense shell pieces,
\begin{equation}
 {{m}_{i}}{{\ddot{R}}_{hs}^i}(t)=4\pi {{{R}_{hs}^i}(t)}^2({{P}_{hs}(t)}-{{P}_{sh}(t=0)}),\label{1} 
\end{equation}	
to assess the hot-spot performance.
Here, ${{m}_{i}}$ represents the mass of the i-th dense shell, ${{P}_{sh}}$ is the shell pressure without perturbation, ${{R}_{hs}^i}$ denotes the hot-spot radius corresponding to the i-th dense shell, ${{P}_{hs}}$ is the shared pressure of the hot spot and $t=0$ represents the onset of deceleration. The kinetic energy of the dense shell at $t=0$ can be expressed as follows, 
\begin{equation}
 {{E}_{k}(t=0)}=\sum\nolimits_{i=1,2,...,N}{m_i{{\dot{{R}}}_{hs}(t=0)}}^2,\label{2} 
\end{equation}	
where \(m_i=\frac{1}{2N}{{\rho }_{0}^i}\delta {{R}_{sh}^i}4\pi {{R}_{hs}(t=0)^2}\), $R_{hs}=R_{hs}^i$ and ${\rho }_{0}^i\delta {R}_{sh}^i$ represents the areal density of the i-th dense shell.
The model inputs consist of five quantities \(v_{{imp}}={\dot{{R}}}_{hs}(t=0)\), \(P_{{hs}}(t=0)-{{P}_{sh}}(t=0)\), \(\sum\nolimits_{i=1,2,...,N}{m_i}\), $R_{hs}(t=0)$ and the area weighted harmonic mean (WHM) of areal density at stagnation. The shell is the plasma between \(R_{{CD}}\) and \(R_p\). The shell velocity \(v_{{imp}}\) is derived from kinetic energy conservation and is equal for all pistons. \(P_{{hs}}\) and \(P_{{sh}}\) are volume‑weighted. The area weighted harmonic mean of areal density is defined as
\begin{equation}
 \mathcal{M}= \frac{\langle 1/(\rho R) \rangle^{-1}}{\langle \rho R \rangle},   
\end{equation}
where \(\langle \bullet \rangle = \int \bullet \, dA / \int dA\), with \(dA\) an element of hot‑spot surface area. $\mathcal{M}$ and the dominant perturbation \(Y_l^0\) are used to jointly constrain the distribution of \(m_i\).

\begin{figure}[h]
\begin{minipage}{0.5\textwidth}
\includegraphics[width=0.93\linewidth]{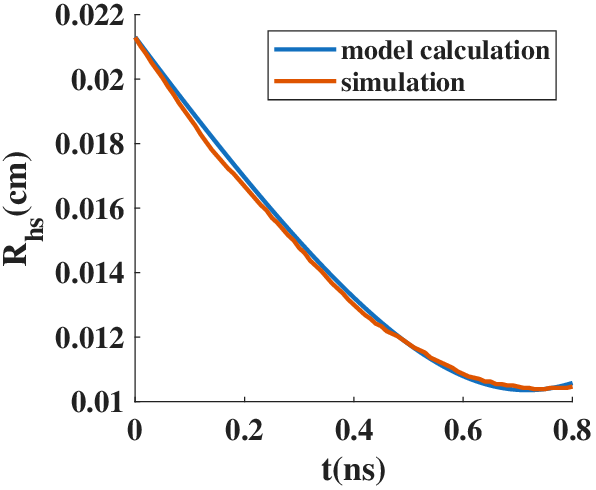}%
\end{minipage}

\caption{\label{FIG. 7}The temporal evolution of hot-spot radius $R_{hs}$ in ideal spherical implosion. The piston model calculation is consistent with FLASH simulations.} 
\end{figure}

In ideal spherical implosions, the piston model, initialized with $\mathcal{M}=1$, yields calculations that match simulations in Fig. \ref{FIG. 7}. Here, \(P_{hs}(t=0)-P_{sh}(t=0)\) is not directly extracted from simulations. It requires fine‑tuning between $[P_{hs}(t=0)-P_{sh}(t=0),P_{hs}(t=0)]$ to match \(R_{hs}\) at stagnation. This fine-tuning compensates for the model's simplified treatment of isobaric hot spot, analogous to adjusting the cold shell entropy \cite{atzeni1984inertial}. The agreement validates that the classical isobaric model adequately captures spherical implosion dynamics. Using these validated parameters, we then apply $\mathcal{M}\neq 1$ to the piston to investigate the influence of hydrodynamic instabilities.

\subsection{The degradation mechanisms of hot‑spot performance with mid‑to-high modes}
To reveal the degradation mechanisms of mid‑to-high modes, we first validate that the isobaric‑adiabatic piston model accurately captures the effects of low‑mode asymmetries. We then compare this validated model, which includes only \(W_{PdV}\), with simulations to isolate the contribution of \(W_{{cond}}\). The comparison shows that neglecting \(W_{{cond}}\) leads to the overestimated pressure and underestimated volume.

\begin{figure}[h]
\begin{minipage}{0.5\textwidth}
\includegraphics[width=0.93\linewidth]{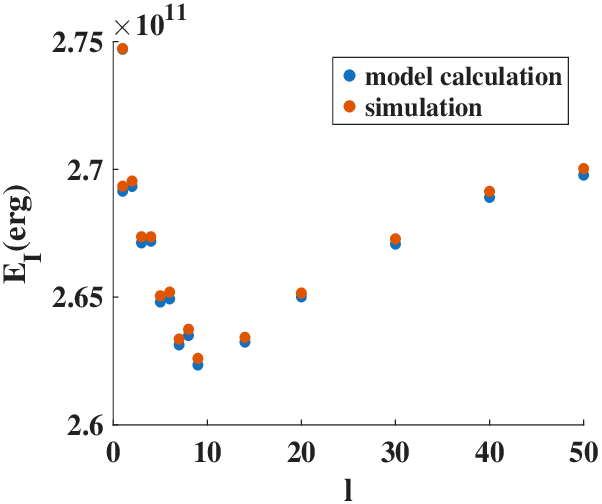}%
\end{minipage}
\caption{\label{FIG. 8} The stagnated hot‑spot internal energy \(E_I\) under the influence of various modes with constant \(\mathcal{A}\). In the piston model, \(\mathcal{M}\) is calibrated to match simulations, ensuring a consistent conversion ratio of shell kinetic energy to \(E_I\).} 
\end{figure}

\begin{figure}[h]
\begin{minipage}{0.47\textwidth}
(a)\includegraphics[width=0.93\linewidth]{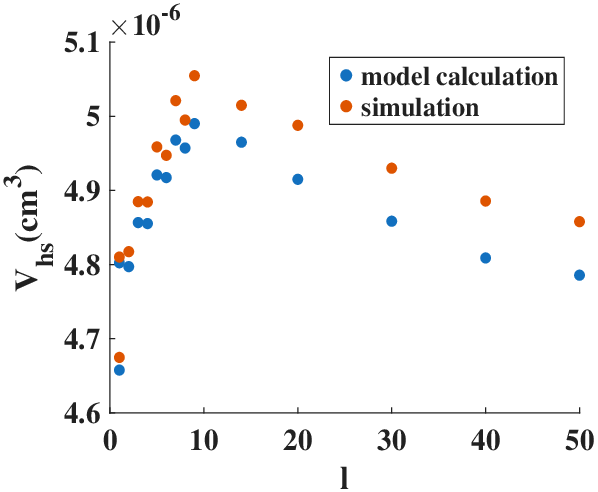}%
 \end{minipage}
\begin{minipage}{0.47\textwidth}
(b)\includegraphics[width=0.93\linewidth]{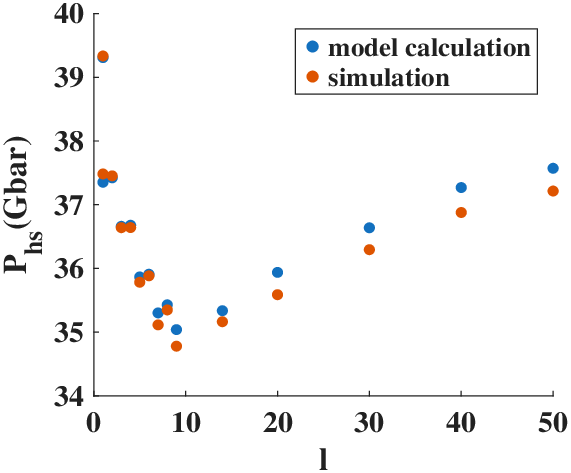}%
 \end{minipage}
 \caption{\label{FIG. 9}Evolution of hot‑spot volume (a) and pressure (b) with perturbation modes of constant amplitude at $t=0$. The
model overestimates pressure and underestimates volume.} 
 
 \end{figure}

Fig. \ref{FIG. 8} shows the stagnated hot‑spot internal energy \(E_I\) under single‑mode perturbations with constant \(\mathcal{A}\) imposed at \(t=0\). Owing to the competition between mode‑dependent perturbation growth and ablation stabilization driven by thermal conduction, \(E_I\) first decreases and then increases with increasing mode number \(l\). To evaluate the influence of hydrodynamic instabilities, we employ the piston model, where the weighted harmonic mean of the areal density $\mathcal{M}$ is tuned to reproduce the simulated \(E_I\). This tuning ensures that the conversion efficiency from shell kinetic energy to $E_{I}$ is consistent with that of simulations. If instabilities degrade hot‑spot performance via \(W_{PdV}\), the stagnated pressure and volume predicted by the model should agree with those from simulations. For example, for \(l \leq 8\), the model accurately captures the increased volume and decreased pressure induced by $W_{PdV}$, as shown in Fig. \ref{FIG. 9}.

If instabilities degrade hot‑spot performance via other mechanisms, the stagnated pressure and volume predicted by the piston model should deviate from those obtained in simulations. As shown in Fig. \ref{FIG. 9}, the validated piston model overestimates pressure and underestimates volume, indicating that additional physical processes are at play. This discrepancy arises from the neglected internal non‑uniformities, which redistribute internal energy via (\(W_{{cond}}\)). Conventional theory holds that \(W_{{cond}}\) does not directly affect hot‑spot pressure, because any lost heat is presumably restored by the inflowing mass. In the hot spot, however, the Peclet number is \(P_e = U\lambda/\alpha \ll 1\), where \(U\), \(\lambda\), and \(\alpha\) are the flow velocity, perturbation wavelength, and the coefficient of thermal diffusivity, respectively. Under such conditions, thermal conduction dominates over convection. Consequently, before the outer mass inflow can restore the pressure in the mixing region, that region has already become isobaric via theermal conduction, resulting in a pressure lower than that in the absence of instabilities. Moreover, for the \(l=50\) mode, whose perturbations are already ablation‑truncated at \(R_{Te}\), both \(W_{{cond}}\) at \(R_{Te}\) and \(W_{PdV}\) at \(R_{CD}\) lead to an enlarged hot‑spot volume and a reduced pressure, further indicating that \(W_{{cond}}\) may degrade performance even when perturbations are nearly stabilized by ablation.

 \(W_{{cond}}\) reducing hot‑spot pressure is applicable across different implosion schemes. This corresponds to a state where $E_{I}$ is the same as that with degradation only induced by $W_{PdV}$, but the pressure and volume differ. It is not captured by any existing models and can be included in a 1D simulations. In future work, we will quantify the mixing region and then describe the thermal conduction process within it using a one‑dimensional thermal source relaxation mixing model \cite{liu2025one}. This approach will naturally incorporate the effects of \(W_{{cond}}\) on hot‑spot pressure and neutron yield under varying Peclet numbers and hot‑spot temperatures.




\section{Conclusions}
In conclusion, for the multi‑scale hydrodynamic instabilities in ignited designs, we propose an hot-spot isobaric criterion,
 which directly links the isobaric assumption to the evolution of perturbations. This criterion states that the isobaric condition remains valid as long as the areal‑density perturbations do not exceed the ratio of the stagnated shell thickness to the coasting distance. The most dangerous mode, \(l=12\), emerges from a balance between perturbation growth and ablation stabilization driven by thermal conduction. When the Peclet number is much less than unity, thermal conduction dominates over convection. Consequently, for mid‑to‑high modes, the hot spot becomes isobaric through thermal conduction before the outer mass inflow can restore the lost heat. Consequently, neglecting thermal conduction overestimates pressure and underestimates volume. These findings not only deepen our understanding of how mid‑to‑high modes degrade hot‑spot performance, but also suggest that thermal conduction losses may degrade performance even when perturbations are nearly stabilized by ablation. Importantly, this mechanism can be captured by one‑dimensional simulations. In future work, we will quantify the mixing region and describe the thermal conduction process within it using a one‑dimensional thermal‑source relaxation mixing model.

\begin{acknowledgments}
This work is supported by Science Challenge Project (Grant No. TZ2025014), and National Natural Science Foundation of China (Grant No. 12375242).
\end{acknowledgments}
\section*{Author Declarations}
\subsection*{Conflict of Interest}
The authors have no conflicts to disclose.
\section*{Data Availability Statement}
Data available on request from the authors.

\section{Appendix}
 The radiation field and target in the 1D MULTI simulation are as follows. The target consists of three concentric layers. The central hot-spot is DT fuel, with a thickness of 983 $\mu \text{m}$ and a density of 6 m$\text{g/cm}^3$. Surrounding this central region is a layer of cryogenic DT fuel, with a thickness of 66 $\mu \text{m}$ and a density of 0.25 $\text{g/cm}^3$. The DT fuel is indicated in red and modeled using the ideal-gas equation of state. The outermost layer is the high-density carbon (HDC) ablator, with a thickness of 79.5 $\mu \text{m}$ and a density of 3.32 $\text{g/cm}^3$. The HDC is shown in blue and described by the Sesame equation of state. The Rosseland and Planck mean opacities are adopted for the radiation transport. The radiation transport is treated with a single-group approximation; while this reduces the ablation efficiency, it does not reduce the ratio of the shell kinetic energy to the hot-spot internal energy. The flux-limiter is 0.06.
 \vspace{100pt}

 \begin{figure}[h]
\begin{minipage}{0.4\textwidth}
\includegraphics[width=0.93\linewidth]{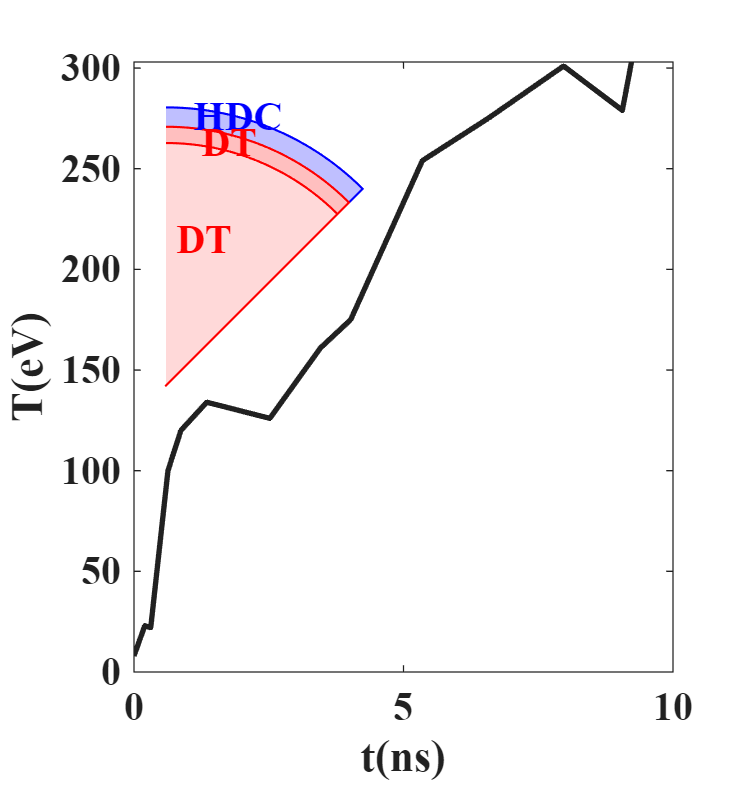}%
 \end{minipage}
 \caption{\label{FIG. appendix}The configurations of radiation and target in the 1D MULTI simulations.} 
 
 \end{figure}

\bibliography{main}

\end{document}